\documentstyle[art10,hip-100a,epsf]{article}
\begin{document}
\begin{center}
{\bf INSTITUT~F\"{U}R~KERNPHYSIK,~UNIVERSIT\"{A}T~FRANKFURT}\\
D-60486 Frankfurt, August--Euler--Strasse 6, Germany
\end{center}

\hfill IKF--HENPG/3--96
\vspace{2cm}

\begin{center}
{\Large {\bf
Strange and Non-Strange (Anti-)Baryon Production at 200 GeV per Nucleon
}}
\end{center}


\vspace{2cm}
\begin{center}
Dieter R\"ohrich 

\vspace{0.5cm}
\noindent
Institut f\"ur Kernphysik, Universit\"at Frankfurt, 
Frankfurt, Germany\\
E--mail address: roehrich@ikf.physik.uni--frankfurt.de

\vspace{5cm}
\noindent
{\it Talk given at STRANGENESS'96, Budapest, Hungary, May 15--17 1996} \\
{\it to be published in the special issue of Heavy Ion Physics}

\end{center}

\vfill

\newpage
%
%
\cimo \setcounter{page}{1} \thispagestyle{empty} \hskip -15pt  \noindent
{\Large\bf
Strange and Non-Strange (Anti-)Baryon Production at 200 GeV per Nucleon
}\\[3mm] 
\def\rightmark{Strange and Non-Strange (Anti-)Baryon Production at 200 GeV/nucl.}\def\leftmark{Dieter R\"ohrich}
\hspace*{6.327mm}\begin{minipage}[t]{12.0213cm}{\large\lineskip .75em
Dieter R\"ohrich \\
for the NA35 Collaboration
}\\[2.812mm] 
\hspace*{-8pt} ~~Institut f\"ur Kernphysik, University of Frankfurt, \\
D--60486 Frankfurt, Germany\\[0.2ex]
\\[4.218mm]{\it
Received nn Month Year (to be given by the editors)
}\\[5.624mm]\noindent
{\bf Abstract.} 
Rapidity distributions of net hyperons ($\Lambda - \overline{\Lambda}$) 
are compared to distributions of participant protons ($p - \overline{p}$).
Strangeness production (mean multiplicities of produced 
$\Lambda/\Sigma^0$ hyperons and $\langle K + \overline{K} \rangle$) 
in central nucleus-nucleus collisions is shown for different 
collision systems at different energies.  
An enhanced production of $\overline{\Lambda}$ compared to 
$\overline{p}$ is observed at 200 GeV per nucleon.

\end{minipage}

\section{Introduction}

Information about baryons and anti-baryons in the hadronic final state
is important for the understanding of the reaction
dynamics of ultrarelativistic nucleus--nucleus collisions.
Experiment NA35 at the SPS measured baryons, antibaryons and
mesons for various collisions systems
at 200 GeV per nucleon \cite{Bam89,Bar90,Bae94,Al:94,Pbar}.  
Data on $\Lambda$/$\overline{\Lambda}$ hyperons can shed light on
various aspects of ultrarelativistic nucleus--nucleus collisions.
By comparing net hyperon ($\Lambda - \overline{\Lambda}$) rapidity
distributions to distributions of participant protons ($p - \overline{p}$)
in minimum bias p+A and central A+A collisions the process of
nuclear stopping can be studied.

Stangeness production is a potential signal for a Quark Gluon Plasma
created in central nucleus--nucleus collisions \cite{Shury,Koch}.
NA35 data on mean multiplicities of strange particles and non-strange  
mesons in comparison with results of other experiments at different energies
allow a systematic study of strangeness production.  
The mean multiplicity of produced
$\Lambda/\Sigma^0$ hyperons and $\langle K + \overline{K} \rangle$
relative to the pion yield is a good measure for the total 
strangeness ($s\overline{s}$) production relative to non-strange mesons. 
This ratio shows an increase in central A+A collisions
compared to nucleon--nucleon interaction. This strangeness enhancement
is observed at Dubna 
(p$_{LAB}$ $\approx$ 4.5 A$\cdot$GeV/c), 
as well as at BNL AGS ($\approx $ 15 A$\cdot$GeV/c) 
and CERN SPS energies ($\approx$ 200 A$\cdot$GeV/c).

The study of antibaryons, such as $\overline{\Lambda}$
and $\overline{p}$, near midrapidity
in central nucleus-nucleus
collisions may shed light on the mechanisms of antiquark production.
The ratio $\overline{\Lambda}$/$\overline{p}$ is
expected to be a good measure for strange quark production as
compared to
the production of non-strange quarks, because the valence antiquark
content of
the $\overline{\Lambda}$ is $\overline{u} \overline{d}
\overline{s}$ and that of the $\overline{p}$ is $\overline{u}
\overline{d} \overline{u}$. This ratio reflects
the yield of $\overline{s}$-quarks relative to that of
non-strange light quarks $\overline{q}$. Furthermore, the
antiquarks of these antibaryons are newly created so that their
distributions
should not directly reflect the distributions of the valence
quarks of the incoming nuclei.

\section{Net Hyperon Rapidity Distributions}

Participant baryons are the final state of those incoming nucleons,
which participate in the interaction. 
Some of the nucleons are
converted into hyperons still carrying two quarks of the
participant
nucleon; their contribution concerning stopping must be also
considered.
The rapidity distributions of ($p-\overline{p}$) and
(${\Lambda-\overline{\Lambda}}$)  for various collision systems
at 200 GeV per nucleon are shown in Figs.~1 and 2. 

\vspace*{10.6cm}
\includegraphics{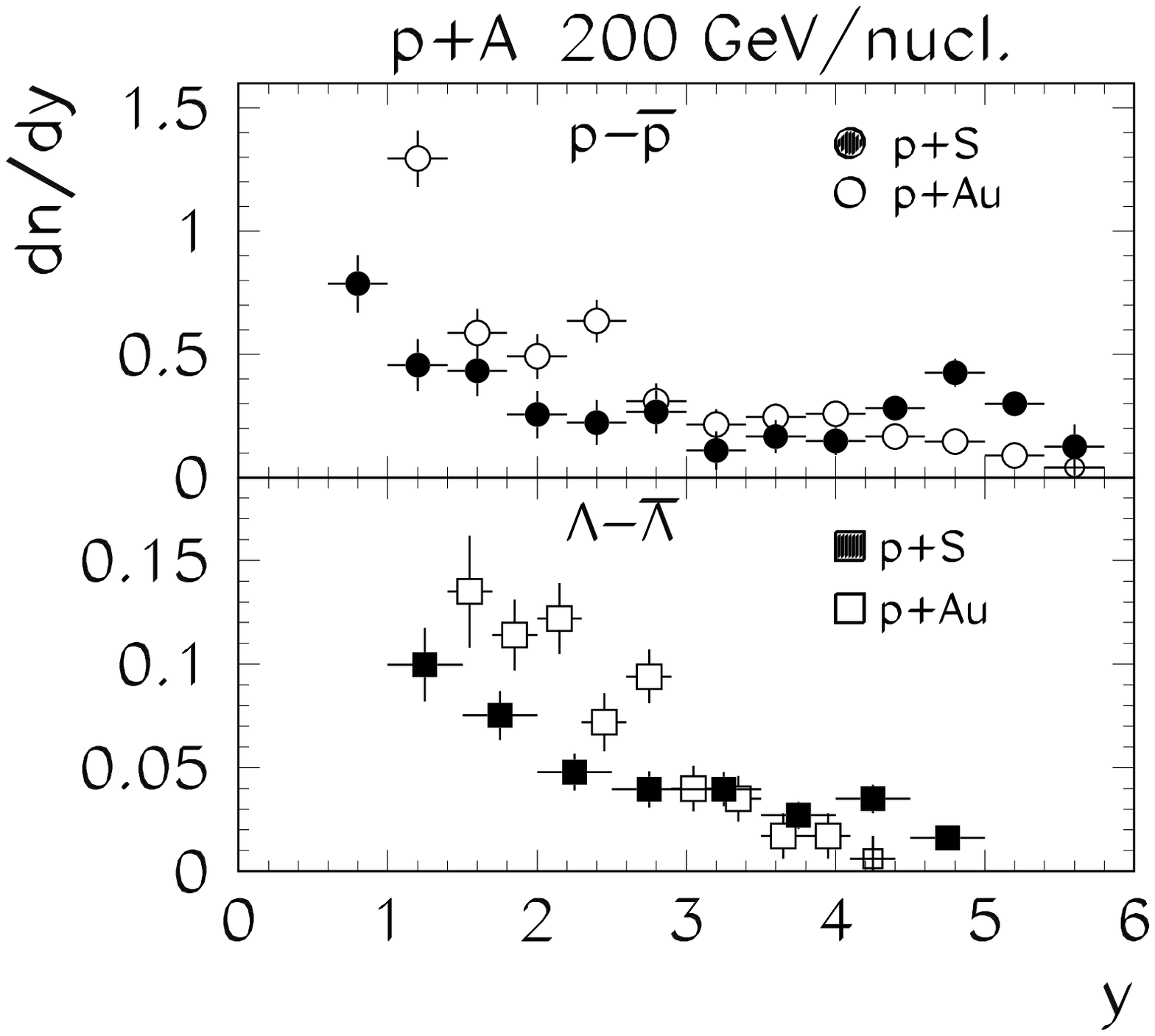}
\vskip -70pt
\begin{minipage}[t]{10cm}
\noindent \bf Fig. 1.  \rm
Rapidity distributions of participant protons $p - \overline{p}$ (top) and
net hyperons $\Lambda - \overline{\Lambda}$ (bottom) in minimum bias p+S and p+Au 
interactions at 200 GeV per nucleon. 
\end{minipage}
\vskip 4truemm

The rapidity distributions of participant protons 
($0.8<y<6.0$) for minimum bias
p+S and p+Au interactions are shown in Fig.~1 (top). While the
rapidity densities around midrapidity for both reactions 
are not very different,
clear differences are observed for rapidities below 1.2 and larger than 4.4.
In p+Au collisions more target nucleons participate in the reactions
and are therefore shifted by about one unit of rapidity. For the projectile
the gold nucleus looks black, i.e. the probability of the projectile
to traverse the target nucleus without loosing any or little energy is
small.
For p+S interactions on the other hand, a clear projectile fragmentation
peak at rapidity 4.7 is visible.
To investigate the effect of changing the nuclear thickness of the target in
heavy ion interactions,
rapidity distributions of participant protons were measured for
central S+S, S+Ag and S+Au collisions. The data are shown in Fig.~2 (top).
The S+S and S+Ag data were measured 
at $0.2<y<3.0$ and
S+Au data at $2.6<y<6.0$.
For the symmetric system S+S,
the data points reflected at $y_{c.m.}$ are also shown.
The shape of the participant proton rapidity distribution changes from
being relatively flat for the
light symmetric system (S+S) to a shape for the
asymmetric systems which monotonically decreases with rapidity.

\vspace*{10.6cm}
\includegraphics{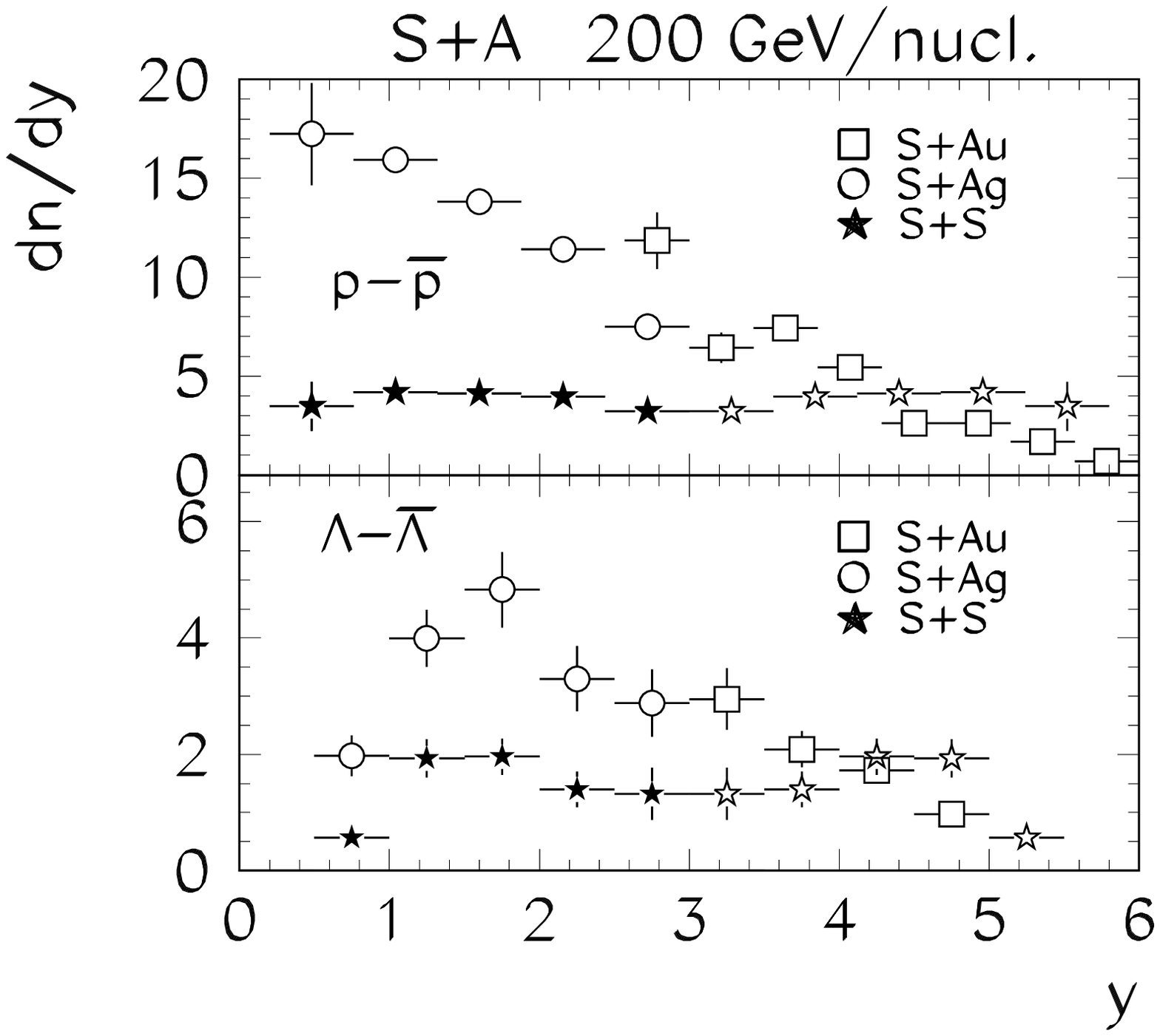}
\vskip -70pt
\begin{minipage}[t]{10cm}
\noindent \bf Fig. 2.  \rm
Rapidity distributions of participant protons $p - \overline{p}$ (top) and
net hyperons $\Lambda - \overline{\Lambda}$ (bottom) in central S+S, S+Ag and
S+Au collisions at 200 GeV per nucleon. 
\end{minipage}
\vskip 4truemm

The rapidity distributions of net hyperons
$\Lambda-\overline{\Lambda}$ for minimum bias p+S ($1.0<y<5.0$) and
p+Au ($1.4<y<4.4$) collisions are displayed in Fig.~1 (bottom),
those for central collisions of S+S ($0.5<y<3.0$), S+Ag ($0.5<y<3.0$), 
and S+Au ($3.0<y<5.0$) are displayed in
Fig.~2 (bottom). The trends in the rapidity
distributions for net hyperons as a function of the target nucleus 
are similar to
those for net protons, but the distributions are clearly compressed
along the rapidity scale, 
they are pushed towards midrapidity by about half a unit
of rapidity due to the energy of $\approx 0.5$ GeV that is needed to produce
a hyperon in a pp-collision.

\section{Strangeness Production}

The measurement of strange particle multiplicities in central
S+A and Pb+Pb collisions by NA35 and NA49 \cite{Veszter,Marg}
allow for a systematic study of strangeness production 
as a function of 
the number of participant nucleons, $\langle N_P \rangle$, 
and as a function of the collision energy \cite{Ga:95a,Ga:96}.
The A+A results are compared with the corresponding results for
all inelastic nucleon--nucleon (N+N) interactions.
In the case of N+N interactions the number of participant nucleons
is taken to be 2.

The total production of strangeness relative to pion production
is studied using the ratio \cite{Al:94}:
\begin{equation}
E_S = \frac { \langle \Lambda \rangle + \langle K + 
\overline{K} \rangle }
{ \langle \pi \rangle },
\end{equation}
where $\langle \Lambda \rangle$ is the mean multiplicity of produced
$\Lambda/\Sigma^0$ hyperons and $\langle K + \overline{K} \rangle$
is the mean multiplicity of kaons and antikaons.
The dependence of the $E_S$ ratio on $\langle N_P \rangle$ is
shown in Fig.~3 for three different collision energies.

\vspace*{12.5cm}
\includegraphics{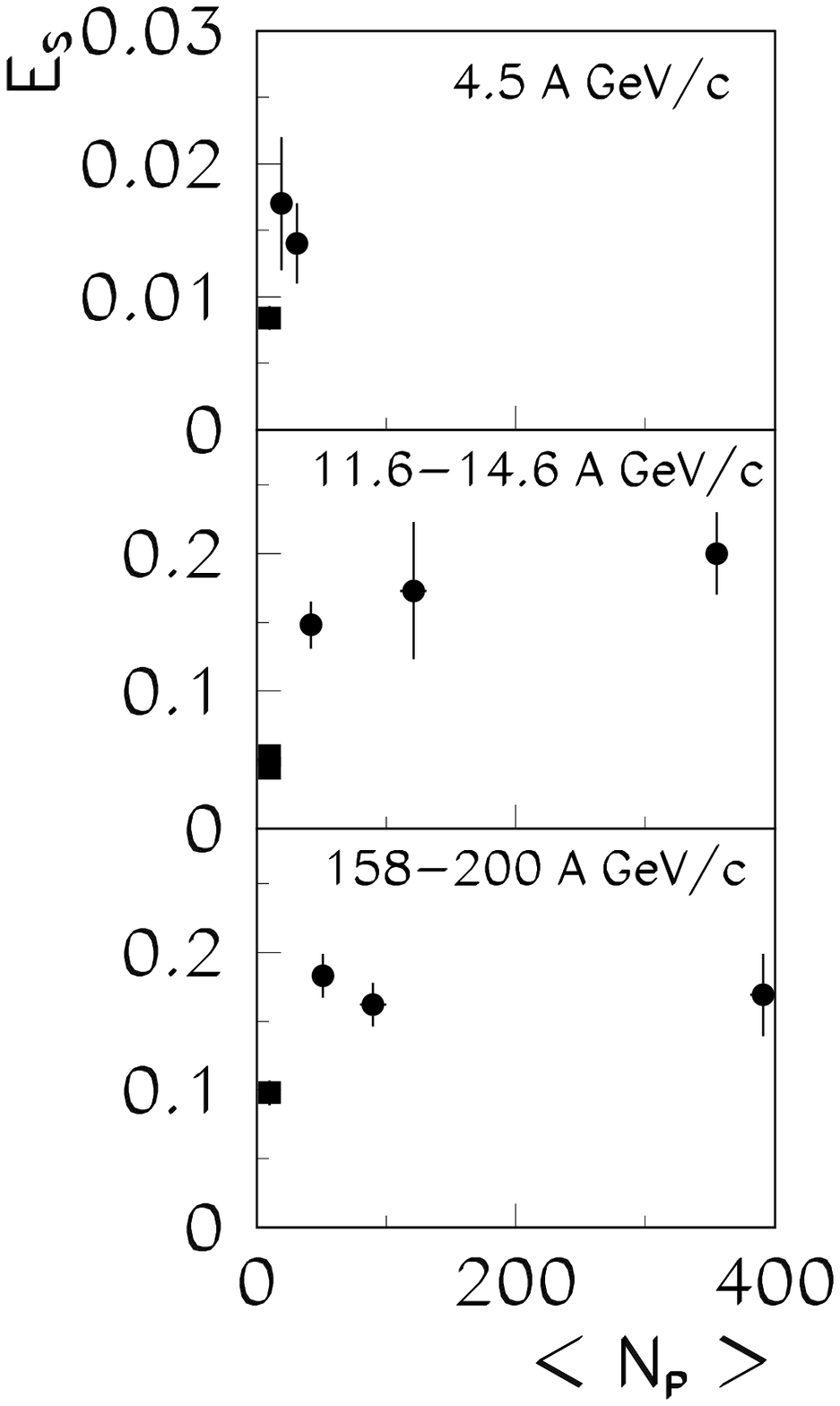}
\vskip -70pt
\begin{minipage}[t]{10cm}
\noindent \bf Fig. 3.  \rm
The dependence of the $E_S$ ratio
on $\langle N_P \rangle$ at three different collisions energies
(p$_{LAB}$ = 4.5 (top), 11.6--14.6 (middle) and 158--200 (bottom) A$\cdot$GeV/c.
\end{minipage}
\vskip 4truemm

The results for central A+A collisions \cite{Ma:96} are shown together with
the results for N+N interactions.
There is a significant increase in the relative strangeness production
(measured by the $E_S$ ratio) at all studied collision energies 
when going from N+N interactions to
central A+A collisions.
This increase is called {\bf strangeness enhancement}.
The relative strangeness production seems to saturate at sufficiently large
values for $\langle N_P \rangle$.
The saturation effect can not be established at 4.5 A$\cdot$GeV/c 
as the results for collisions of heavy nuclei do not exist at this
collision energy.
The effect of strangeness enhancement is largest at AGS-energies 
($\approx$ factor of 4) as shown in Fig.~4.

\vspace*{9.0cm}
\includegraphics{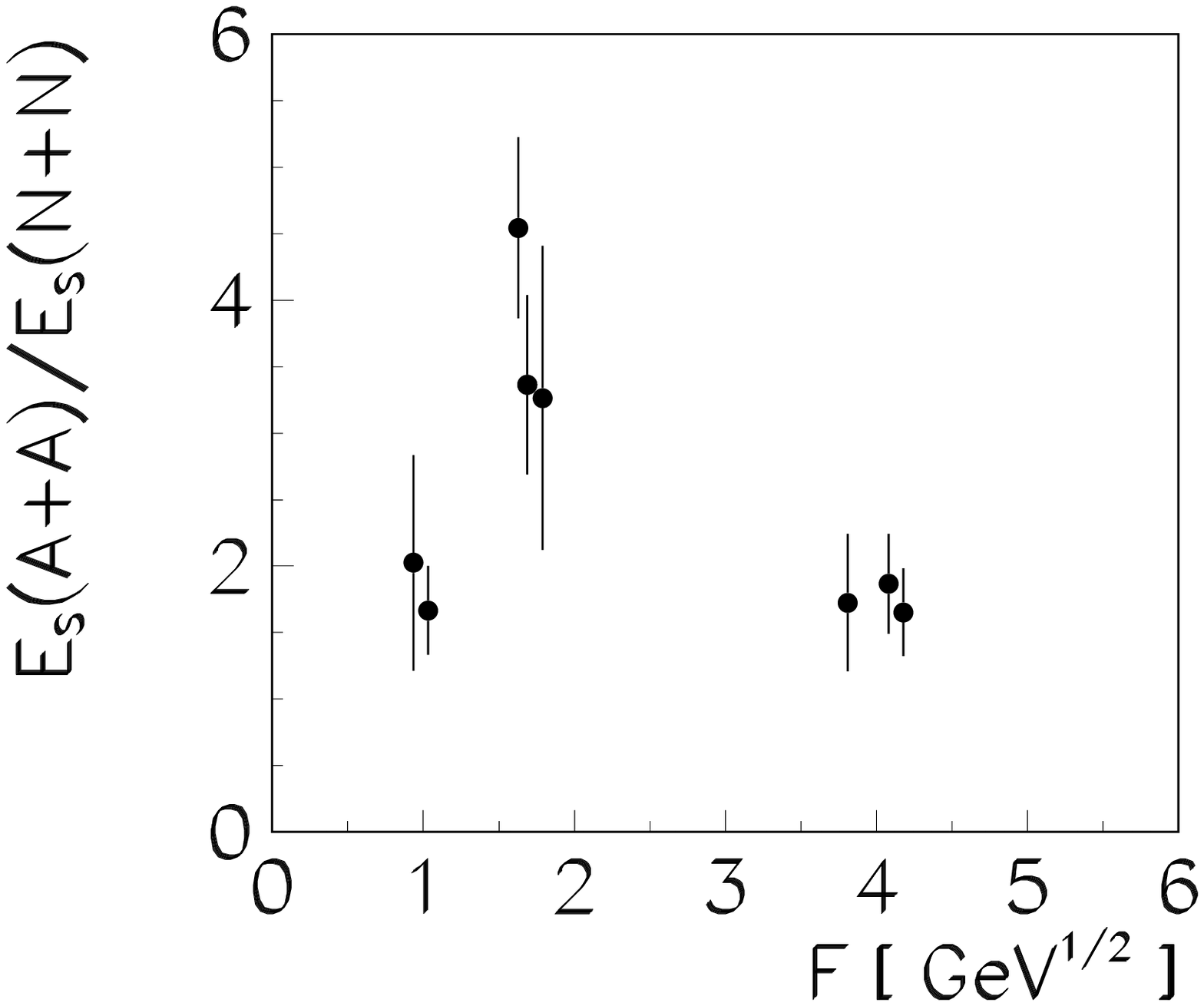}
\vskip -70pt
\begin{minipage}[t]{10cm}
\noindent \bf Fig. 4.  \rm
The dependence of the $E_S$ ratio (see Eq.~1) in central 
A+A collisions relative 
to the ratio in N+N interactions on the
collision energy measured by the Fermi energy variable, $F$ (see
Eq. 2).
\end{minipage}
\vskip 4truemm

The collision energy dependence is studied  using Fermi energy
variable \cite{Fe:50,La:53}
\begin{equation}
F = \frac {(\sqrt{s}_{NN} - 2 m_N)^{3/4} } { \sqrt{s}_{NN}^{1/4} },
\end{equation}
where $\sqrt{s}_{NN}$ is the c.m. energy for a nucleon--nucleon pair and
$m_N$ is the mass of the nucleon.
For a detailed discussion of this analysis see the contribution of 
Marek Ga\'zdzicki to this conference \cite{Marek}.

The collision energy dependence of the $E_S$ ratio 
for central A+A collisions is shown in Fig.~5 (bottom),
the results for N+N interactions are shown in Fig.~5 (top).
A monotonic increase of $E_S$ for N+N interactions between Dubna energy 
(p$_{LAB}$ = 4.5 A$\cdot$GeV/c) and CERN SPS energy 
(p$_{LAB}$ = 200 A$\cdot$GeV/c)
is observed;
in the range from 15 A$\cdot$GeV/c to 200 A$\cdot$GeV/c the $E_S$ ratio
increases by a factor of about 2.
A qualitatively different energy dependence of the $E_S$ ratio is
observed for central A+A collisions.
The rapid increase of the $E_S$ between Dubna and BNL AGS energies
is followed by a weak change in the $E_S$ between BNL AGS and CERN SPS
collision energies.
For a detailed discussion see the contribution of
Marek Ga\'zdzicki to this conference \cite{Marek}.

\vspace*{11.0cm}
\includegraphics{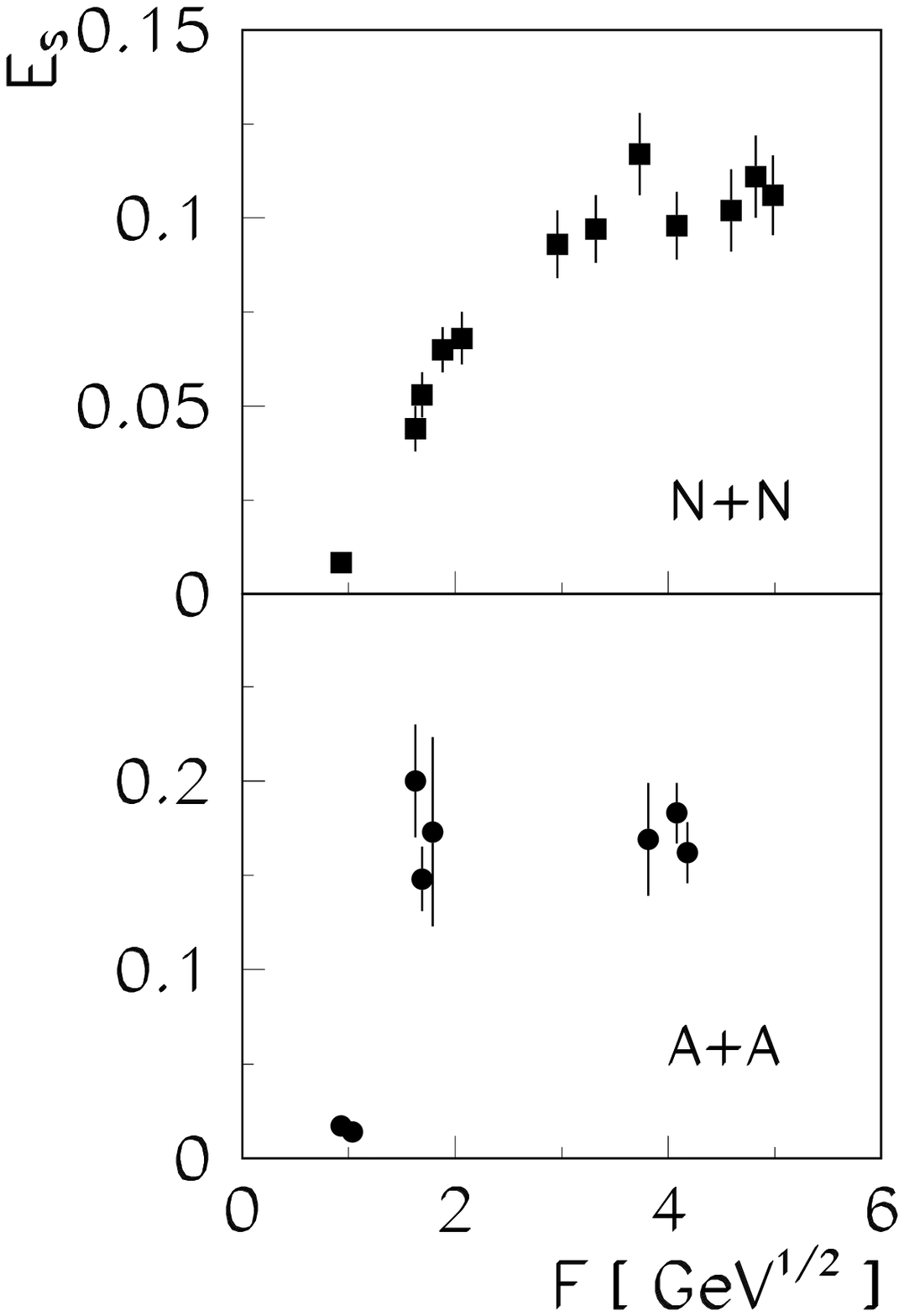}
\vskip -70pt
\begin{minipage}[t]{10cm}
\noindent \bf Fig. 5.  \rm
The dependence of the $E_S$ ratio
for N+N interactions (top) and central A+A collisions (bottom) on the
collision energy measured by the Fermi energy variable, $F$ (see
Eq. 1).
\end{minipage}
\vskip 4truemm

\section{Antibaryon production}

The ratio $\overline{\Lambda}$/$\overline{p}$ provides
information on the production of strange antiquarks compared to
light antiquarks.
Antibaryon
yields in central nucleus-nucleus collisions are expected to exhibit
the subtle
interplay between various partonic and/or hadronic production and
annihilation processes as well as
properties of a possible partonic
equilibration of the system \cite{Sorge_et_al}.
Theoretical consideration of the possible creation of a
Quark Gluon Plasma in ultrarelativistic nuclear collisions
\cite{Shury}
has indicated that strange/antistrange quark pairs might be copiously
produced
resulting in an approximate flavour symmetry among light quarks,
namely
up, down and strange \cite{Koch}. Under such conditions the
$\overline{\Lambda}$/$\overline{p}$\,-ratio, which roughly reflects
the $\overline{s}$/$\overline{u}$\,-ratio, should approach
unity (or larger for non-zero baryon densities) \cite{Koch},
far exceeding the value of
$\sim$ 0.25 observed in proton-proton collisions.

The $\overline{\Lambda}$/$\overline{p}$\,-ratio near midrapidity
in proton-proton, minimum bias proton-nu\-cle\-us and central
nucleus-nucleus collisions is shown in Fig.~6 as a function
of the rapidity density at midrapidity of negatively charged
hadrons $h^-$ \cite{Bam89,Bar90,Al:94,Pbar}. The ratio
increases from a value of 0.25 for p+p (and p+A) collisions
to a value of approximately 1.4 for central S+A collisions

\vspace*{8.6cm}
\includegraphics{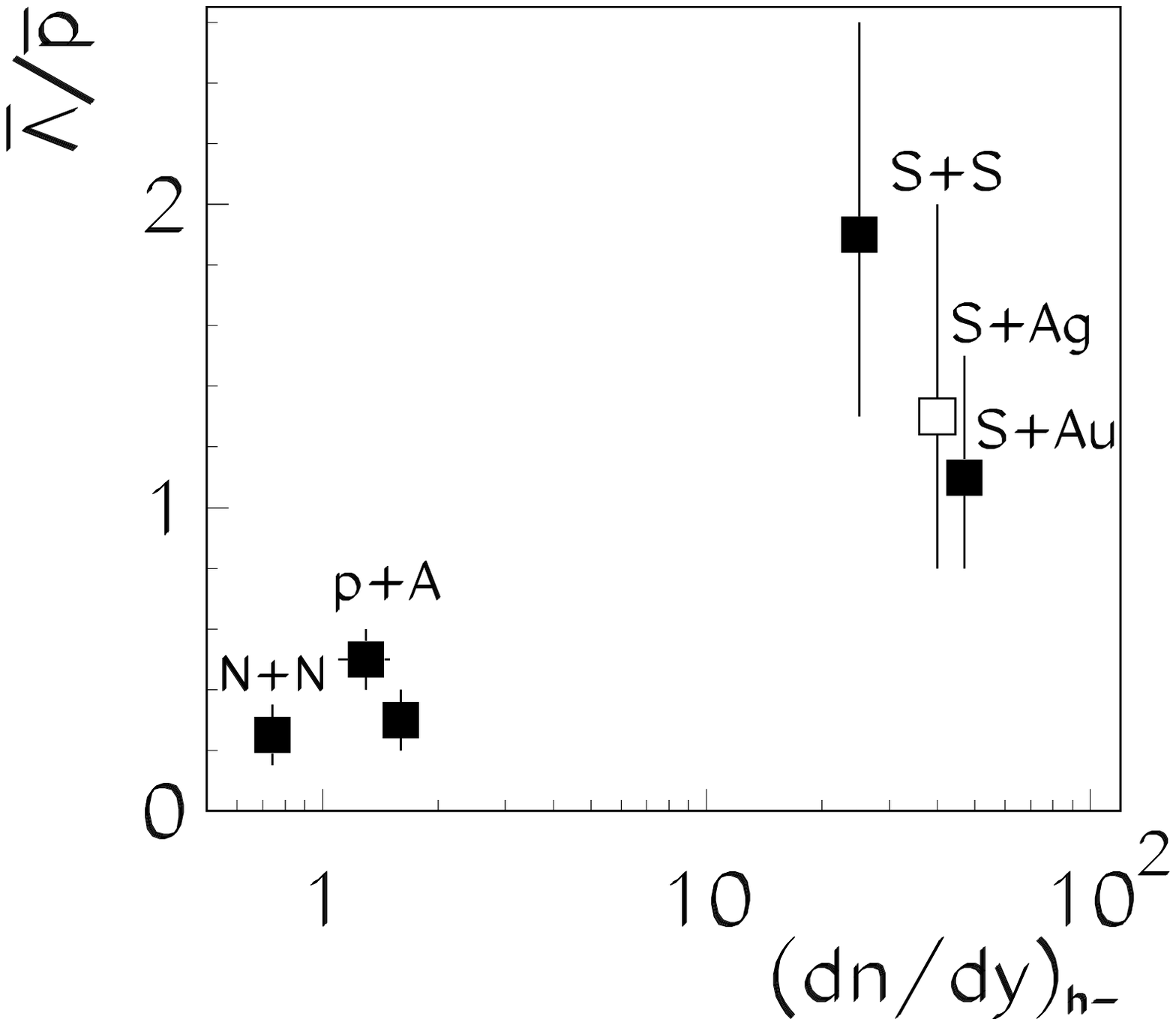}
\vskip -70pt
\begin{minipage}[t]{10cm}
\noindent \bf Fig. 6.  \rm
{$\overline{\Lambda}$/$\overline{p}$\,-ratio near
midrapidity
in proton-proton, minimum bias proton-nucleus and central
nucleus-nucleus collisions ($3<y<4$) at 200 GeV per nucleon as a function of
the rapidity density of negatively charged hadrons
at midrapidity. The result for S+Ag
(open symbol) is not the result of a direct measurement, the 
$\overline{\Lambda}$ yield was interpolated between S+S and S+Au collisions.}
\end{minipage}
\vskip 4truemm

\vspace*{8.4cm}
\includegraphics{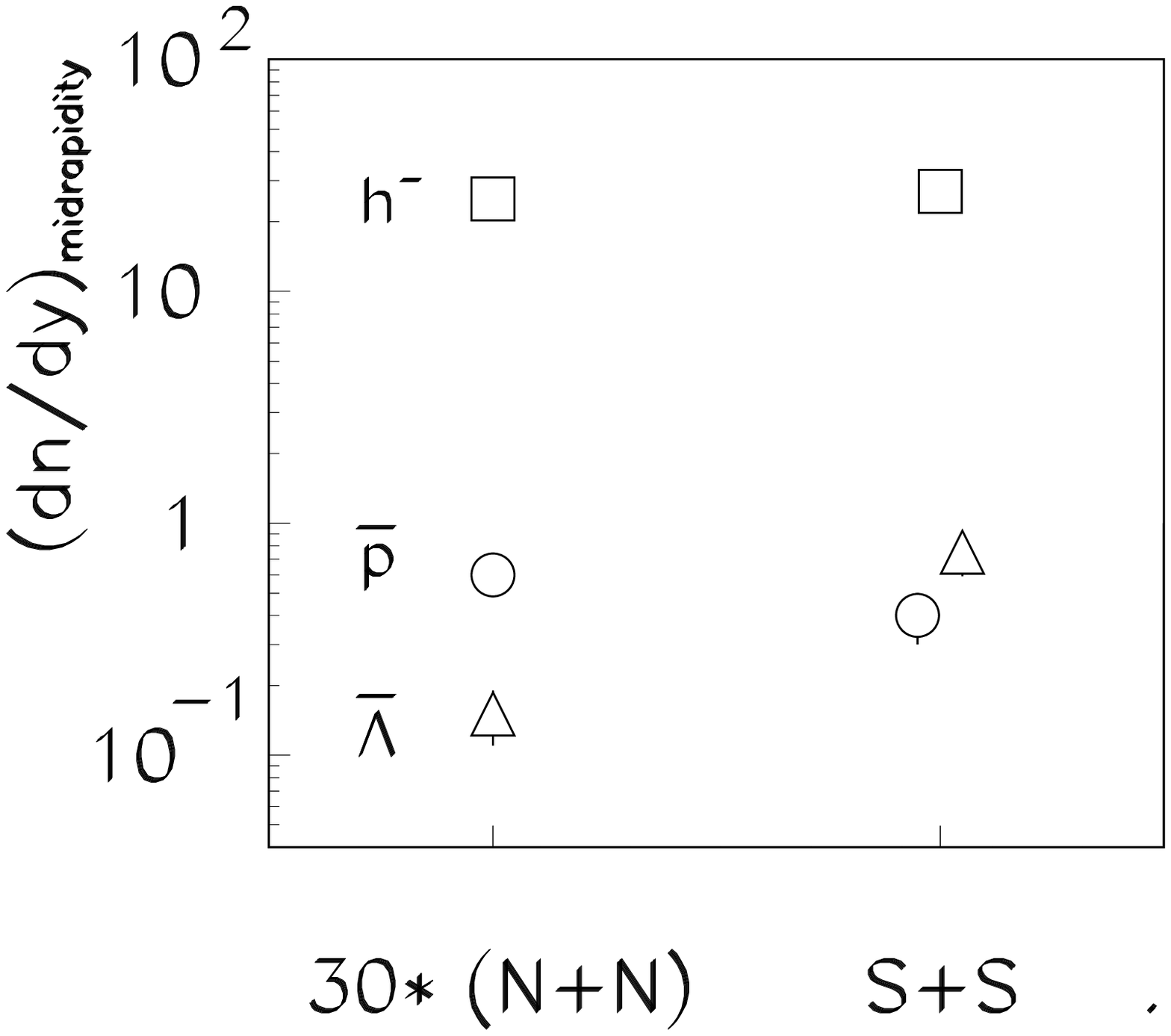}
\vskip -70pt
\begin{minipage}[t]{10cm}
\noindent \bf Fig. 7.  \rm
$\overline{\Lambda}$ and $\overline{p}$ yields near
midrapidity
in nucleon-nucleon and central
S+S collisions at 200 GeV per nucleon. The N+N data 
are scaled by the ratio of the total $h^-$-multiplicities.
\end{minipage}
\vskip 4truemm

The increase in the ratio can be studied in more detail for the
S+S system. If rapidity densities at midrapidity in nucleon-nucleon
collisions are scaled by the ratio of $h^-$-multiplicities in
central S+S and in nucleon-nucleon collisions (Fig.~7),
the $\overline{p}$ production decreases by about 30 $\%\/$  and
the $\overline{\Lambda}$ abundance at midrapidity is strongly
enhanced by a factor of 5.

\section{Summary}

The trends in the rapidity distributions of net hyperons
$\Lambda-\overline{\Lambda}$ 
as a function of projectile and target nucleus are similar to
those for net protons, but the distributions are 
different.
The experimenal data on strangeness production indicate
a saturation of strangeness production with the
number of participant nucleons
and a change in the collision energy dependence occuring between
15 A$\cdot$GeV/c and 200 A$\cdot$GeV/c.
The ratio of strange antibaryon $\overline{\Lambda}$ to the non-strange
antibaryon $\overline{p}$ in central A+A collisions at 200 GeV per 
nucleon is larger than
one and therefore significantly larger than the ratio in nucleon--nucleon
interactions.

\vfill\eject

\end{document}